\documentclass[reprint,aps,prl,twocolumn,superscriptaddress]{revtex4-1}
\usepackage{graphicx}
\usepackage{dcolumn}
\usepackage{bm}
\usepackage{amsmath}
\usepackage[english]{babel}
\usepackage[utf8]{inputenc}
\usepackage{color}
\begin{document}

\title{Light effective hole mass in undoped Ge/SiGe quantum wells}

\author{M. Lodari}
\affiliation{QuTech and Kavli Institute of Nanoscience, Delft University of Technology, PO Box 5046, 2600 GA Delft, The Netherlands}
\author{A. Tosato}
\affiliation{QuTech and Kavli Institute of Nanoscience, Delft University of Technology, PO Box 5046, 2600 GA Delft, The Netherlands}
\author{D. Sabbagh}
\affiliation{QuTech and Kavli Institute of Nanoscience, Delft University of Technology, PO Box 5046, 2600 GA Delft, The Netherlands}
\author{M. A. Schubert}
\affiliation{IHP -- Leibniz-Institut für innovative Mikroelektronik, Im Technologiepark 25, 15236 Frankfurt, Germany}
\author{G. Capellini}
\affiliation{IHP -- Leibniz-Institut für innovative Mikroelektronik, Im Technologiepark 25, 15236 Frankfurt, Germany}
\affiliation{Dipartimento di Scienze, Universit\`a degli studi Roma Tre, Viale Marconi 446, 00146 Roma, Italy}
\author{A. Sammak}
\affiliation{QuTech and Netherlands Organisation for Applied Scientific Research (TNO), Stieltjesweg 1, 2628 CK Delft, The Netherlands}
\author{M. Veldhorst}
\affiliation{QuTech and Kavli Institute of Nanoscience, Delft University of Technology, PO Box 5046, 2600 GA Delft, The Netherlands}
\author{G. Scappucci}
\email{g.scappucci@tudelft.nl}
\affiliation{QuTech and Kavli Institute of Nanoscience, Delft University of Technology, PO Box 5046, 2600 GA Delft, The Netherlands}

\date{\today}

\begin{abstract}
We report density-dependent effective hole mass measurements in undoped germanium quantum wells. We are able to span a large range of densities ($2.0-11\times10^{11}$ cm$^{-2}$) in top-gated field effect transistors by positioning the strained buried Ge channel at different depths of 12 and 44 nm from the surface. From the thermal damping of the amplitude of Shubnikov-de Haas oscillations, we measure a light mass of 0.061$m_e$ at a density of $2.2\times10^{11}$ cm$^{-2}$. We confirm the theoretically predicted dependence of increasing mass with density and by extrapolation we find an effective mass of $\sim0.05m_e$ at zero density, the lightest effective mass for a planar platform that demonstrated spin qubits in quantum dots. 
\end{abstract}

\maketitle
Holes are rapidly emerging as a promising candidate for semiconductor quantum computing.\cite{maurand_cmos_2016,watzinger2018germanium,hendrickx2019fast} In particular, holes in germanium (Ge) bear favorable properties for quantum operation, such as strong spin-orbit coupling enabling electric driving without the need of microscopic objects,\cite{maurand_cmos_2016,watzinger2018germanium,hendrickx2019fast} large excited state splitting energies to isolate the qubit states,\cite{terrazos_qubits_2018} and ohmic contacts to virtually all metals for hybrid superconducting-semiconducting research\cite{dimoulas_fermi-level_2006,katsaros_hybrid_2010,hendrickx_ballistic_2019,vigneau_germanium_2019,xiang2006ge}. Furthermore, undoped planar Ge quantum wells with hole mobilities $\mu>5\times 10^{5}$ cm$^2$/Vs were recently developed\cite{sammak_shallow_2019} and shown to support quantum dots\cite{hendrickx_gate-controlled_2018,hardy_single_2019} and single and two qubit logic,\cite{hendrickx2019fast} providing scope to scale up the number of qubits. 

Holes in strained Ge/SiGe quantum wells have the attractive property of a light effective mass parallel to the Ge well interface.\cite{winkler1996theory,schaffler1997high,terrazos_qubits_2018} This property is highly desirable for spin qubits since it provides large energy level spacing in quantum dots, allowing to relax lithographic fabrication requirements and enhance tunnel rates. The light effective hole mass is due to the compressive strain in the quantum well, which splits the heavy hole and light hole bands and induces a mass inversion, i.e. the topmost band develops a lighter mass than the lower-lying band.\cite{schaffler1997high} An effective hole mass of 0.05$m_e$ was recently predicted\cite{terrazos_qubits_2018} for Ge/Si$_{1-x}$Ge$_{x}$ heterostructures with alloy concentrations $x\sim0.75$, corresponding to strain levels accessible experimentally.

Previous studies in modulation doped Ge/SiGe heterostructures showed, indeed, a very light effective mass of 0.055$m_e$\cite{morrison2017electronic}, measured in Hall-bar devices aligned with the ${<}110{>}$ crystallographic direction and further reduced  to 0.035$m_e$ for the ${<}100{>}$ direction. The nonparabolicity effects of the valence bands\cite{irisawa2003hole,rossner2003effective,sawano2006magnetotransport,rossner2006effective} tend to increase the effective mass, with smaller values expected at lower hole densities $p$ due to the decreasing of the associated Fermi vector.

\begin{figure}[!ht]
	\includegraphics[width=85mm]{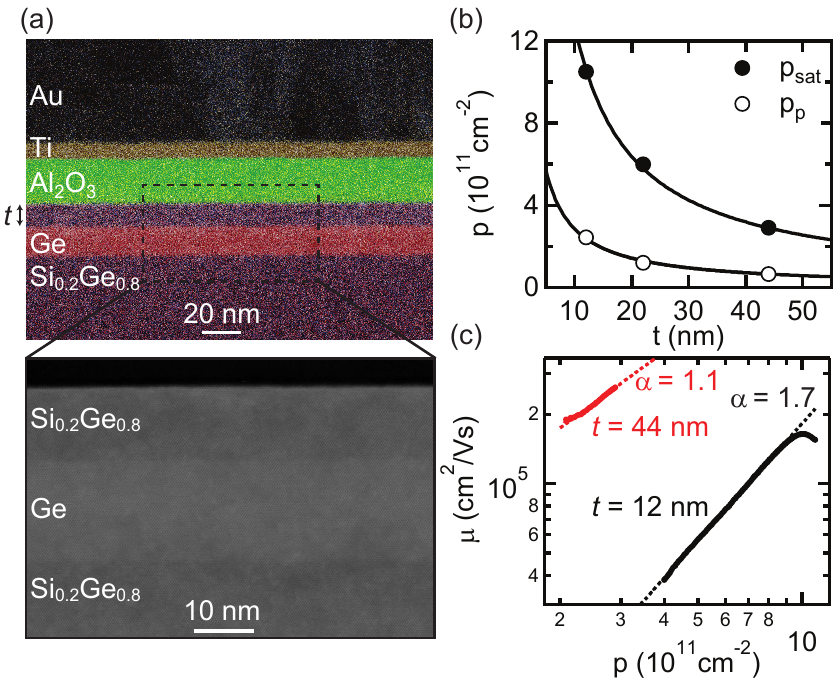}%
	\caption{(a) STEM-EDX and TEM of a Ge/SiGe heterostructure field effect transistor with the quantum well positioned 12 nm under the gate stack. (b) Saturation $p_{sat}$ and percolation density $p_p$ as a function of the position of the quantum well $t$. Curves are fits to a $\sim1/t$ dependence. Data for $t = 22$ nm are extracted from Ref \cite{sammak_shallow_2019}. (c) Density-dependent mobility $\mu(p)$ and power law fit.}
\label{fig:MAT}
\end{figure}

Modulation doping, however, exhibits impurities that are a source for charge noise, disorder, gate leakage, and device instability at low temperature\cite{borselli2011pauli}. Therefore, undoped Ge/SiGe quantum wells are preferable for quantum dot fabrication.\cite{hendrickx_gate-controlled_2018} The transport properties of undoped Ge/SiGe quantum wells are relatively unexplored and effective mass measurements have shown so far conflicting results. In Ref. \cite{laroche_magneto-transport_2016} a rather large effective mass of 0.105$m_e$ was reported at a low density of $1\times10^{11}$ cm$^{-2}$. Furthermore, no clear dependence of the effective mass with density could be extracted in the investigated range from $\sim0.6\times10^{11}$ cm$^{-2}$ to $\sim1.4\times10^{11}$ cm$^{-2}$. In Ref. \cite{hardy2018singlePREPRINT,hardy_single_2019}, instead, a lighter mass was reported with a nearly constant value of 0.08$m_e$ over the measured density range ($\sim1-4\times10^{11}$ cm$^{-2}$).

In this Letter we reconcile experiments with theoretical expectations and provide evidence that the effective hole mass in low-disorder undoped Ge/SiGe decreases towards lower densities. We measure a minimum effective mass value of 0.061$m_e$ at a density of $2.2\times10^{11}$ cm$^{-2}$, which extrapolates to $0.048(1)m_e$ at zero density. This makes strained Ge/SiGe the planar platform with the lightest effective mass for spin qubit devices.

The undoped Ge/SiGe heterostructures are grown by reduced-pressure chemical vapor deposition and comprise a Si$_{0.2}$Ge$_{0.8}$ virtual substrate, a 16-nm-thick Ge quantum well (in-plane compressive strain of -0.63\%) and a Si$_{0.2}$Ge$_{0.8}$ barrier. Two heterostructures of different barrier thickness are considered ($t$ = 12, 44 nm). Hall-bar shaped heterostructure field effect transistors (H-FET) are fabricated aligned along the ${<}110{>}$ direction using a low-thermal budget process which features platinum-germanosilicide ohmic contacts and an Al$_2$O$_3$/Ti/Au gate stack. Magnetotransport characterization of the devices is performed at temperature $T=1.7-10$ K using standard four-probe low-frequency lock-in techniques. A negative bias applied to the gate induces a two-dimensional hole gas and controls the carrier density in the quantum well. Details of the heterostructure growth, device fabrication and operation, and magnetotransport measurements are reported in Ref. \cite{sammak_shallow_2019}.  

Figure \ref{fig:MAT}(a) shows scanning transmission electron microscopy with energy dispersive X-ray (STEM-EDX) analysis of the shallow Ge quantum well ($t$ = 12 nm) under the gate stack. These images highlight the overall quality of the strained Ge H-FET. A uniform quantum well of constant thickness is obtained, and sharp interfaces are observed between the quantum well and the barrier and between the barrier and the dielectric layer.

The position of the quantum well determines the range of accessible density $p$ in these Ge H-FETs. At a given $t$, the density range extends from the percolation threshold density $p_p$ (Fig. \ref{fig:MAT}(b), open circles) to the saturation density $p_{sat}$ (Fig. \ref{fig:MAT}(b) , solid circles). Saturation of carriers in the quantum well is achieved at high gate bias when the Fermi level aligns with the valence band edge at the dielectric/SiGe interface.\cite{lu2011upper} We observe a $p_{sat}\sim1/t$ dependence, as expected from Poisson's equation, indicating that charges in the system are in the equilibrium state.\cite{su_effects_2017} The percolation threshold density represents the critical density for establishing metallic conduction in the channel. This is extracted by fitting the density-dependent conductivity in the low density regime to percolation theory,\cite{tracy2009observation,kim2017annealing} as applied in Ref. \cite{sammak_shallow_2019} to Ge H-FETs. We observe a $\sim1/t$ dependence, expected for long-range scattering from remote impurities at the dielectric/semiconductor interface.\cite{gold1991metal,gold2011metal}

Figure \ref{fig:MAT}(c) shows the density-dependent mobility $\mu$ at $T = 1.7$ K. The observed power law dependence $\mu\sim p^\alpha$ is characterized by an exponent $\alpha$ of 1.6 and 1.1 in the shallow (black line, $t = 12$ nm) and deeper quantum well (red line, $t = 44$ nm), respectively. The $\alpha$ values confirm that the mobility is limited by scattering from the dielectric/semiconductor interface, as previously observed in Si/SiGe and Ge/SiGe H-FETs.\cite{laroche2015scattering,mi2015magnetotransport,su_effects_2017,sammak_shallow_2019} Despite the close proximity to the dielectric interface, the shallower quantum well has a remarkable peak mobility of $1.64\times 10^{5}$ cm$^2$/Vs at $p = 1.05\times10^{12}$ cm$^{-2}$, 2.4$\times$ larger than previous reports for quantum wells positioned at a similar distance from the surface.\cite{su_effects_2017} At higher density the mobility starts to drop, possibly due to occupation of the second subband or to different scattering mechanisms becoming dominant. The deeper quantum well ($t = 44$ nm) has a higher  mobility of $2.6\times 10^{5}$ cm$^2$/Vs at a much lower density of  $2.9\times10^{11}$ cm$^{-2}$, as expected due to the larger separation from the scattering impurities. We therefore find, by using Ge H-FETs with different $t$, that high values of mobility are achieved over a large range of density, making these devices well suited for Shubnikov-de Haas (SdH) measurements of the density-dependent effective mass.

\begin{figure}[!ht]
	\includegraphics[width=85mm]{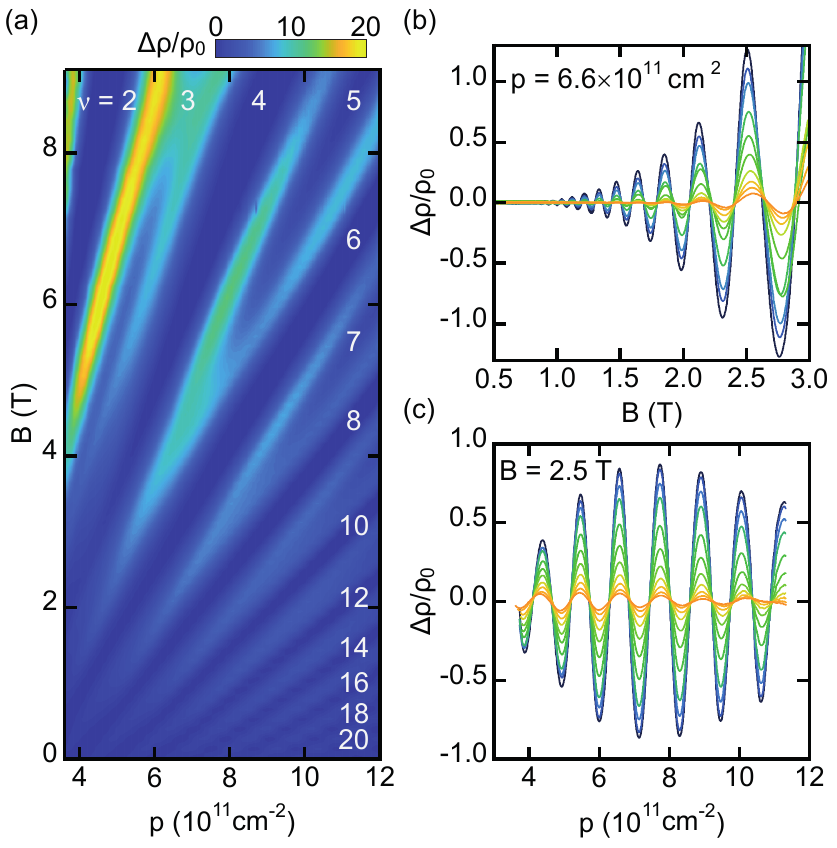}%
	\caption{(a) Fan diagram at $T=1.7$ K showing the magnetoresistance normalized to the zero field value $\Delta\rho_{xx}/\rho_{0}$ as a function of $B$ and $p$ for the sample with $t=12$ nm. Filling factors $\nu$ assigned from quantum Hall effect are indicated. (b) Temperature dependence ($T=1.7-10$ K) of $\Delta\rho_{xx}/\rho_{0}$ at fixed $p$, variable $B$ and (c) at fixed $p$, variable $B$. Data in (b) and (c) are plotted after polynomial background subtraction.}
\label{fig:fan}
\end{figure}

In Fig. \ref{fig:fan}(a) we show a Landau fan diagram for the shallow quantum well ($t$ = 12 nm). This is obtained by plotting the oscillatory component of the magnetoresistivity $\Delta\rho_{xx}/\rho_{0}=(\rho_{xx}(B)-\rho_{0})/\rho_{0}$ at $T=1.7$ K as a function of out-of-plane magnetic field $B$ and carrier density $p$, obtained from the low-field Hall data. Shubnikov-de Haas oscillations fan out towards higher field and density, with Zeeman spin splitting visible at odd filling factors $\nu$. Temperature dependence of the oscillation amplitudes are shown in Fig.~\ref{fig:fan}(b) and (c) after a polynomial background subtraction. Fig.~\ref{fig:fan}(b) shows the cross-section of the fan diagram at fixed density, obtained by keeping the gate voltage constant while sweeping the magnetic field. Alternatively, the density is swept at a fixed magnetic field (Fig.~\ref{fig:fan}(c)). Both data sets allow the estimate of the effective mass with a better insight into the dependence on $B$ and $p$. 
The effective mass $m^*$ is obtained by fitting the thermal damping of the SdH oscillations by using the expression\cite{de1993effective}\begin{equation}
\dfrac{\Delta\rho/\rho_{0}(T)}{\Delta\rho/\rho_{0} (T_{0})}=\dfrac{T\sinh{(\beta T_{0})}}{T_{0}\sinh{(\beta T)}} ,  
\end{equation}
where $\beta=\dfrac{2\pi k_{B} m^* m_{e}}{{\hbar}eB}$, $k_{B}$ is the Boltzmann constant, $\hbar$ is the Plank constant, $e$ is the electron charge and $T_0=1.7$ K is the coldest temperature at which the oscillations were measured. 

\begin{figure}[!ht]
	\includegraphics[width=85mm]{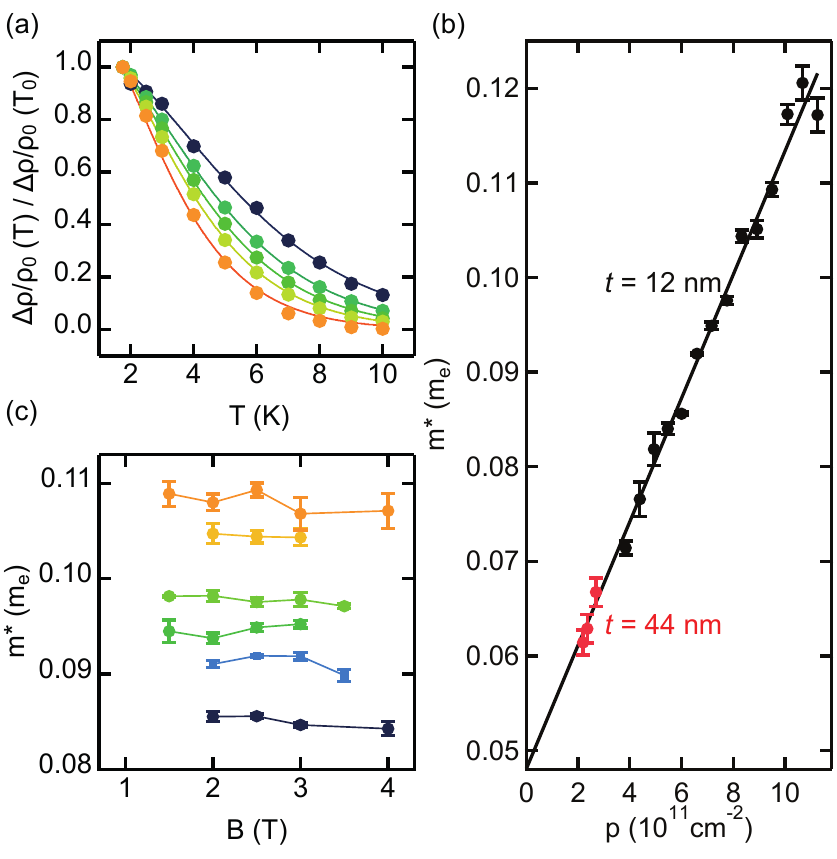}%
	\caption{(a)$\Delta\rho_{xx}/\rho_{0}$ (solid circles) as a function of $T$, normalized at $\Delta\rho_{xx}/\rho_{0}(T_0)$, with $T_0 = 1.7$ K. Different colors correspond to different densities from $3.84\times10^{11}$ cm$^{-2}$ (dark blue circles) to $10.66\times10^{11}$ cm$^{-2}$ (orange circles). Colored lines are theoretical fits used to extract $m^{*}$ as a function of density. (b) Density dependent $m^*$ and linear extrapolation to zero density. The data at $t=44$ nm was obtained by sweeping the magnetic field at a fixed density, while those at $t=12$ nm were obtained by sweeping the density at fixed magnetic field. (c) Magnetic field dependent $m^*$ at different densities from $\sim5.8\times10^{11}$ cm$^{-2}$ (dark-blue solid circles) to $\sim9.5\times10^{11}$ cm$^{-2}$ (orange solid circles).}
\label{fig:mass}
\end{figure}

In Fig.~\ref{fig:mass}(a) experimental data and theoretical fitting are shown for different densities at a fixed magnetic field $B = 2.5$ T. The resulting $m^*$ values are reported as a function of the correspondent density $p$ in Fig.~\ref{fig:mass}(b) for both quantum wells. We observe a strong increasing mass with density, which nearly doubles over the range of investigated densities. The magnetic field dependence of the mass (Fig.~\ref{fig:mass}(c)) is rather weak in the investigated range ($B\leq 4$ T), which is limitied to SdH oscillations before Zeeman splitting. The linear density-dependent effective mass extrapolates to $m^* = 0.048(1)m_e$ at zero density. This value is in agreement with the predicted theoretical value calculated from the density of states at the $\Gamma$ point,\cite{terrazos_qubits_2018} reconciling theory and experiments. 

In summary, we have measured the effective hole mass over a large range of densities in  high-mobility undoped Ge/SiGe quantum wells. The obtained values (0.061$m_e$), extrapolated to 0.048(1)$m_e$ at zero density, are the lightest effective mass reported for a planar platform that demonstrated spin qubits in quantum dots. These results position planar germanium as a promising material towards the development of spin and hybrid quantum technologies.

\section{Acknowledgments}
We acknowledge support through a
FOM Projectruimte of the Foundation for Fundamental Research on Matter (FOM), associated with the Netherlands Organisation for Scientific Research (NWO).

\providecommand{\noopsort}[1]{}\providecommand{\singleletter}[1]{#1}%

\end{document}